\documentclass[twocolumn,showpacs,preprintnumbers,amsmath,amssymb]{revtex4}

\usepackage{graphicx}
\usepackage{amsmath}
\usepackage{color}

\begin{document}
\title{Lattice Interferometer for Ultra-Cold Atoms}
\author{Mikkel F. Andersen}

\affiliation{Atomic Physics Division, National Institute of Standards and Technology, Gaithersburg, Maryland 20899-8424, USA}
\affiliation{Jack Dodd Center for Quantum Technology, Department of Physics, University of Otago, New Zealand}
\author{Tycho Sleator}
\affiliation{New York University, Department of Physics, 4 Washington Place, New York, NY 10003} 
\begin{abstract}

We demonstrate an atomic interferometer based on ultra-cold atoms released from an optical lattice. This technique yields a large improvement in signal to noise over a related interferometer previously demonstrated. The interferometer involves diffraction of the atoms using a pulsed optical lattice. For short pulses a simple analytical theory predicts the expected signal. We investigate the interferometer for both short pulses and longer pulses where the analytical theory break down. Longer pulses can improve the precision and signal size. For specific pulse lengths we observe a coherent signal at times that differs greatly from what is expected from the short pulse model. The interferometric signal also reveals information about the dynamics of the atoms in the lattice. We investigate the application of the interferometer for a measurement of $h/m_A$ that together with other well known constants constitutes a measurement of the fine structure constant.

\end{abstract} 
\date{July 7, 2008}
\maketitle

Matter wave interference has intrigued scientists since the early days of quantum mechanics. Still today its fundamental nature is a field of intense research, and beautiful demonstrations and investigations shedding light on this phenomenon have been performed in the past decade. These include interference of ``large'' objects and of bio-molecules \cite{Hackermuller04}, interference of independently prepared particles \cite{Andrews97}, and the origin of quantum mechanical complementarity \cite{Durr98}.
Advances in microfabrication techniques and the development of laser cooling and trapping for neutral atoms has opened up many new possibilities for constructing atomic interferometers \cite{Keith88,Weiss93}. Besides testing the fundamental nature of matter wave interference, atom interferometers play an essential role in many high precision measurements, where the accurate determination of fundamental constants, such as the fine structure constant $\alpha$ and the Newtonian constant of gravity, by several independent means, tests the borders of our understanding \cite{Weiss93, Fixler07,Clade06}. Moreover, precise measurements of quantities such as the local gravitational field hold promise for technological advances in navigation and mineral exploration \cite{McGuirk02}.

In this letter we demonstrate a new type of echo-interferometer
\cite{Weiss93,Keith91,Weitz96,Cahn97,Gupta02,Turlapov05}. It is very simple and uses atoms initially laser cooled and loaded into a one-dimensional (1D) optical lattice potential, released, and later exposed to a pulse of the lattice potential. This single pulse interferometer enhances the signal by more than a factor of four compared to the two pulse interferometer in Ref.~\cite{Cahn97}. It can be used for a precision measurement of the recoil frequency $\omega_{k}=\hbar k^2/(2m_A)$,  where $m_A$ is the atomic mass and $k$ the wave vector of a laser field. This quantity is of great interest for testing fundamental theories \cite{Weiss93,Clade06, Gupta02}. 
The interferometer we present here shares the robustness against vibrations, accelerations, rotations, magnetic field gradients and differences of AC Stark shifts between internal levels of those demonstrated in \cite{Gupta02,Cahn97}, without the systematic errors due to interactions between the atoms the use of a Bose-condensate inflicts in Ref.~\cite{Gupta02}.

We investigate how our interferometer performs when the optical lattice pulse violates the ``short'' pulse or Raman-Nath limit (A regime not previously investigated in related interferometers), and find that a moderate violation of this limit can enhance the performance of the interferometer. We show that the interferometer reveals information on the quantum dynamics of atoms in an optical-lattice potential, and thereby that it holds promise for use in the study of driven one-dimensional systems, a very active field of research in the past decade (see e.g. Refs \cite{Raizen99,Bruchleitner06,Kanem07}). Furthermore we find that for specific pulse lengths, a coherent signal can occur at times that differ from the expected echo time by as much as 10 times the coherence time expected from the initial momentum spread of the atoms. We compare all the results to a simple numerical model and find excellent agreement.

\begin{figure}
\includegraphics[width=8.6cm]{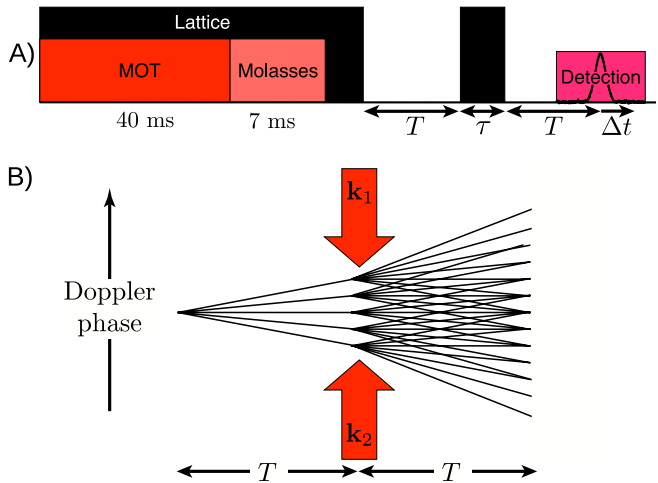}
\caption{A) Time line of the experimental cycle.  B) Doppler phase diagram for atoms released from optical lattice.} 
\label{fig:timeline}
\end{figure}

Figure \ref{fig:timeline} shows a timeline of our experiment, which uses a vapor cell magneto optical trap (MOT) for $^{85}$Rb atoms loaded for 40 ms. An optical molasses stage of 7 ms further cools the atoms and loads them into the optical lattice, which is formed by two vertically polarized horizontally propagating laser beams with wavevectors $\mathbf{k}_1$ and  $\mathbf{k}_2$, an angle of 162 degree between them, and detuned 395 MHz above the $|5S_{{1}/{2}},F=3 \rangle$ to $|5P_{{3}/{2}},F'=4 \rangle$ transition. The lattice laser beams are clipped Gaussian beams with a diameter larger than the MOT cloud, so all the atoms in the MOT are loaded into the lattice. After the molasses stage, which cools the atoms to $\sim \!36 \,\mu$K, the repump laser remains on for 100 $\mu$s to prepare the atoms in the $F=3$ ground state. We control the optical lattice beams using acusto-optic modulators and 10 $\mu s$ after the turn-off of the repump light we abruptly turn off the optical lattice. We denote the time the lattice is turned off $t=0$. We then leave the atoms in darkness for a time $T$ after which we pulse the optical lattice on for a short time $\tau$. At a later time we detect the amplitude of atomic density modulations with period $2 \pi / q $ with $\mathbf{q}= \mathbf{k}_1-\mathbf{k}_2 $ by applying a weak off-resonant optical field along direction $\mathbf{k}_1$ and measuring the amplitude of the field Bragg scattered off the atomic density modulation along the direction $\mathbf{k}_2$ using a heterodyne technique \cite{Cahn97}. 
%

To calculate the expected signal from our interferometer, we assume that the laser-cooled atoms are in thermal equilibrium in the optical lattice, and localized to regions near the potential minima much smaller than an optical wavelength, so the atoms form a periodic density distribution and the potential experienced by them can be approximated by a harmonic oscillator potential. Since the temperature of the atoms is $\sim \!36 \,\mu$K their thermal de Broglie wave coherence length is much shorter than the period of the optical lattice. 
Under these conditions, it can be shown that the atoms in our experiment 
closely approximate an incoherent mixture of states identical to the ones obtained by an atomic plane wave of momentum $\mathbf{q}_0$ impinging on a periodic array of Gaussian transmission functions with width $\sigma$ (the width of the atomic density distribution in a single minimum in the lattice) and period $2\pi/q$:
\begin{equation}
\psi_{\mathbf{q}_0}(\mathbf{x},0) = \sum_{n_1=-\infty}^\infty e^{-n_1^2(q\sigma)^2} e^{i(\mathbf{q}_0 + n_1 \mathbf{q})\cdot\mathbf{x}}\,,
\label{eq1}
\end{equation}
where each state $\psi_{\mathbf{q}_0}$ contributes with a weight given by the momentum ($\hbar \mathbf{q}_0$) distribution of a gas of atoms in thermal equilibrium. 
The experimental signal can be computed by first calculating the signal resulting from the system initially in state $\psi_{\mathbf{q}_0}(\mathbf{x},0)$ and then summing this signal over the distribution of $\mathbf{q}_0$.  
 
After the lattice is turned off at a time $t=0$, each plane wave, $\exp[i(\mathbf{q}_0+n\mathbf{q})\cdot\mathbf{x}]$ in Eq.~(\ref{eq1}) acquires a phase 
$\phi = (\omega_{q_0} + n_1^2\omega_q + n_1 \mathbf{q}\cdot \mathbf{v}_0)\,t$, where $\omega_q = \hbar q^2/2 m$ is the (two-photon) recoil frequency and 
$n_1 \mathbf{q}\cdot \mathbf{v}_0 t$ is the Doppler phase (the component of the phase proportional to the initial atomic velocity $\mathbf{v}_0 =\hbar \mathbf{q}_0/m$). 
The optical lattice pulse, turned on at time $t=T$, diffracts each plane wave into a set of plane waves with wave vectors differing by integer multiples of $\mathbf{q}$ \cite{Ovchinnikov99}.
%
%
%
%
If $\tau$ is so short that atomic motion can be neglected during the pulse (Raman-Nath condition), then no Doppler phase evolution occurs during this time. 
Figure \ref{fig:timeline}B) shows a diagram of the Doppler phase evolution of various amplitudes as a function of time in our interferometer.
%
%
Crossing lines in the diagram occur at times when different momentum states have the same Doppler phase, and atomic fringe patterns are produced at these times.  
%
%
In particular, fringe patterns with period $2\pi/q$ are produced close to times $t_N = (N+1) T$ for positive integer $N$ ($t_N$ are called the echo-times).
%
%
%
A detailed calculation similar to the one in Ref. \cite{Cahn97},
which assumes 
that the interaction during the pulse is given by $H =\frac{1}{2} V_0 \cos(\mathbf{q}\cdot \mathbf{x})$, gives a signal proportional to
\begin{equation}
S(\Delta t) \!=\! 
e^{-(q u \Delta t/2)^2}\! e^{-\frac{1}{2}N^2(q\sigma)^2} \!
J_{\!N+1}\![2\theta \sin(N \omega_q T+\omega_q\Delta t)].
\label{eq2}
\end{equation}
where $u=\sqrt{2 k_B \mathcal{T}/m_A}$, $k_B$ Boltzmann's constant, $\mathcal{T}$ the temperature of the atoms, $\Delta t\equiv t-t_N$, and $\theta =  \tau V_0/(2 \hbar)$ is a pulse area. 



\begin{figure}
\includegraphics[width=8.6cm]{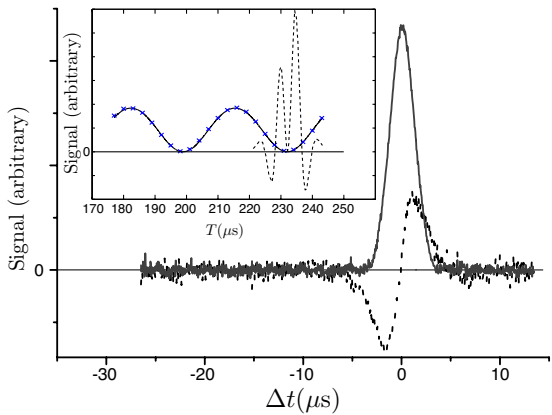}
\caption{Solid curve: Signal from lattice interferometer for $\tau=350$ ns and $T=81\,\mu$s. Dashed curve: Maximal signal we could obtain from the interferometer described in \cite{Cahn97}. Inset:  Peak signal as a function of pulse separation $T$ for small area pulses $\tau=100$ ns ($\times\!$'s).  Peak signal theoretically expected from Eq.~(\ref{eq2}) using measured $V_0$ (solid curve). Dashed curve shows the measured peak signal for $\tau=1.2\,\mu$s, too long for the Raman-Nath condition. The longer pulse yields sharper features in the signal.}
\label{fig:nvso}
\end{figure}

Figure \ref{fig:nvso} shows the signal obtained at around $T = 81\, \mu$s with a pulse duration $\tau = 350$~ns (short pulse approximation still valid). All data shown correspond to the fundamental echo $N=1$. Our temperature estimate is found by fitting Eq.~(\ref{eq2}) to the data.
Note that the signal reaches a maximum at the echo time ($\Delta t=0$). This is in contrast to the signal obtained from the interferometer in \cite{Cahn97} (also shown in Fig.~\ref{fig:nvso} for a similar number of atoms), where no density modulation of the atoms occurs exactly at the echo time, but just before and after. The signals shown in Fig.~\ref{fig:nvso} are the largest we could obtain in the short pulse limit with the laser power and detuning we use. The maximum signal size of our interferometer is more than a factor of four larger than that of Ref.~\cite{Cahn97}, demonstrating an improved signal-to-noise ratio and higher contrast of the atomic density modulation. 
We ascribe this to the fact that in our interferometer the signal is an echo of a density modulation of the atoms, whereas in Ref. \cite{Cahn97} it is a velocity (or phase) modulation, that with time evolves into a density modulation, but also partially dephases due to the thermal velocity spread of the atoms. By laser cooling the atoms into the optical lattice we avoid the large loss of atoms, associated with using e.g. an optical mask \cite{Turlapov05} (the atom optics analog of an absorption grating in light optics) for generation of the atomic density modulation. 

Equation (\ref{eq2}) also shows that the peak signal at the echo time varies periodically as a function of $T$ with period given by the Talbot time $T_T = 2\pi/\omega_q$.  Our interferometer can therefore be used to measure the the Talbot time (or equivalently, the recoil frequency), which together with other well known constants constitutes a measurement of the fine structure constant $\alpha$ \cite{Weiss93,Clade06,Gupta02}.
The inset of Fig.~\ref{fig:nvso} shows the analytical prediction [Eq.~(\ref{eq2})] together with the experimental measurements of the peak signal as a function of $T$ for a pulse length of 100 ns. In the analytical prediction we use $V_0/h=2.0$ MHz determined in a separate measurement of the $\tau$ that yield the first maximum in signal for $T=81\,\mu$s. The overall amplitude was adjusted to fit the data.
By comparing the size of the echo signal for $N = 1, \,2$, and 3, we can use Eq.~(2) to extract the degree of localization $\sigma$ of the atoms in the lattice, and find that $\sigma = 55$ nm.



\begin{figure}
\includegraphics[width=8.6cm]{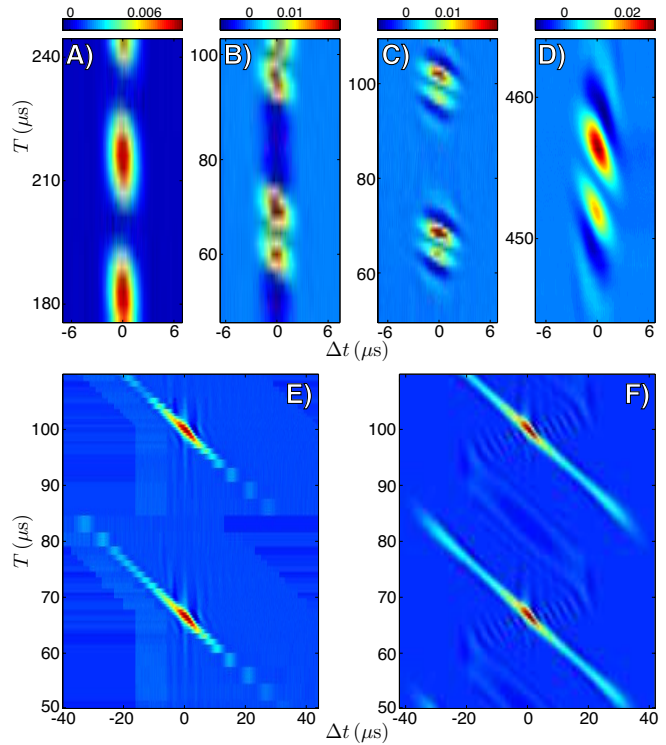}
\caption{Echo signal (color coded) as a function of $\Delta t$ and time of the pulse $T$ (the echo time is $t_1=2T+\tau$). The inset in Fig.~\ref{fig:nvso} is a vertical cross-section through the plots in this figure at the echo time ($\Delta t=0$), and the main data in Fig.~\ref{fig:nvso} is a horizontal cross-section of the data in this figure. A), B), and C): data for pulse durations of $\tau=100$ ns, $600$ ns, and $1.2\,\mu$s. For $\tau=1.2\,\mu$s the signal vanishes for all $T$ around $80\,\mu$s in contrast to the prediction of Eq. \ref{eq2}. D) high resolution data $\tau=1.2\,\mu$s. E) Data for $\tau=3.5\,\mu$s. Features as function of $T$ are no longer as sharp as for $\tau=1.2\,\mu$s. We observe a coherent signal for times that differs highly from the echo time. F) Numerical calculation of the expected signal for the parameters of E).}
\label{fig:all}
\end{figure}

Sharp features in the interferometric signal as a function of $T$ (or equivalently higher frequency components in the signal) improves the precision with which we can determine the Talbot time and recoil frequency \cite{Cataliotti01}. From Eq.~(\ref{eq2}) we see that if we increase $\theta$, more oscillations and sharper features occur in each period when $T$ is scanned, thereby improving the sensitivity of the interferometer. $V_0$ and thereby $\theta$ can be increased by increasing the power in the lattice beams, but the above results are obtained using the maximum laser power we have available. Increasing $\tau$ will also increase $\theta$, but this will eventually lead to a break down of the Raman-Nath condition, and thereby the validity of Eq.~(\ref{eq2}). 
When the Raman-Nath condition is violated the signal is still periodic in $T$ with period $T_T$ independent of pulse duration. Thus the recoil frequency can be determined simply from this period.
We therefore investigate what happens when the pulse length is increased beyond the Raman-Nath limit. In Fig.~\ref{fig:all} we present the echo signal as a function of $T$ and $\Delta t$ for different pulse durations $\tau$.  
For $\tau = 1.2\,\mu$s [Fig.~3C)] we see a clear deviation from the results predicted by Eq.~(\ref{eq2}), namely that the signal vanishes for $T$s around $(n+1/2)T_T/2$ with $n$ an integer, and that the signal is asymmetric around $nT_T/2$. However, the narrow ``dark'' fringe around $nT_T/2$ persists, enabling an accurate determination of $T_T$. Our experimental observation that the narrowest features of the echo signal as a function of $T$, are found for pulse durations between $1\,\mu$s and $2\,\mu$s are not surprising, since the optical lattice imparts maximum momentum into the atoms for durations around $\tau \sim 1/4 \tau_{osc}$ ($\tau_{osc}$ is the oscillation period of an atom close to a potential minimum).
We use the sharp features described above to determine the Talbot time by taking data with high resolution in $T$ for $\tau=1.2\,\mu$s around $T=65 \mu s$ and around $T=455\,\mu$s -differing in $T$ around 6 Talbot times [see Fig.~\ref{fig:all}D)]. From this we obtain a value of $h/m_\mathrm{Rb}=4.6997 \times 10^{-9} \pm 0.0003 \times 10^{-9}$ $m^2/s^2$ \footnote{The uncertainties are one standard deviation combined systematic and statistical.}, 
with $m_{Rb}$ the mass of a $^{85}$Rb atom, in agreement with the value of $h/m_\mathrm{Rb}=4.6994 \times 10^{-9}$ $m^2/s^2$ deduced from \cite{Audi03}. Our ``large'' uncertainty arises from the determination of the angle between the beams, a problem that can be overcome by using counter propagating beams and coupling between optical fibers \cite{Clade06}. 

For $T=3.5\,\mu$s [see Fig. \ref{fig:all}E)] we observe the interesting phenomenon that a coherent signal for certain values of $T$ is observed for times that differ from the echo time ($2T+\tau$) by as much as $40\,\mu$s. This is more than 10 times the decoherence time of few $\mu$s expected from the initial thermal spread of atoms. The nature of the signal also seems to indicate its occurrence is not due to long coherence times, but rather because the dynamics of the atoms during the lattice pulse enables a coherent rephasing at this time. We note that this phenomenon occurs for pulse durations $\tau$ around $\frac{1}{2}\tau_{osc}$. In Fig.~\ref{fig:all}F) we show a 1D numerical calculation of the expected signal for the same parameters as the experimental results in Fig.~\ref{fig:all}E) (no photon scattering is included in the calculation).

\begin{figure}
\includegraphics[width=8.6cm]{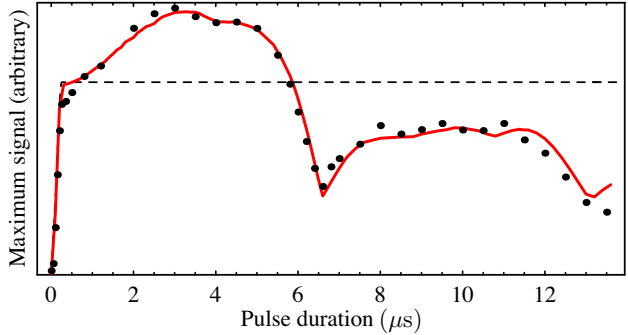}
\caption{Maximum signal as a function of pulse duration.  Dots are the data, and the solid curve is a numerical calculation multiplied by an exponential decay to accommodate for decoherence due to photon scattering. The lifetime of the exponential decay is found by fitting to the data. Dashed line is the maximum signal expected from Eq.~(\ref{eq2}). In contrast to the experimental data and numerical calculations no oscillations are seen. We observe that violating the Raman-Nath condition can increase the contrast of the atomic density modulation.}
\label{fig:dyn}
\end{figure}
To further investigate the dynamics of the atoms in the optical lattice we measured the maximum signal size for a given $\tau$ by scanning $T$. In Fig.~\ref{fig:dyn} we plot this maximum signal as a function of $\tau$. We observe that violating the Raman-Nath limit can improve the contrast of the atomic density modulation since for pulse lengths from 1 to 5 $\mu$s, we observe a larger signal than predicted by Eq.~\ref{eq2} (also shown in Fig \ref{fig:dyn}). In contrast to the prediction of Eq. \ref{eq2} the signal shows damped oscillations with a period of around 6.6 $\mu s$. This period is consistent with $\tau_{osc}$ reflecting a partial revival of the initial state at this time. This effect has been observed previously using a BEC in an optical lattice \cite{Ovchinnikov99}, and the fact that it easily is seen in our data, indicates that our interferometer can be used as a sensitive probe of the quantum dynamics in diffracting structures, including classical chaotic systems such as the $\delta$-kicked rotor \cite{Raizen99}. In Fig.~\ref{fig:dyn} we also show a numerical calculation of the expected maximum signal as a function of $\tau$. To account for decay of coherence due to photon scattering in the optical lattice the numerical calculation shown in Fig.~\ref{fig:dyn} has been multiplied by an exponential decay as function of $\tau$ where the decay rate of $3.5\times 10^4$ s$^{-1}$ is found by fitting to the data. This decay rate is smaller than the average photon scattering rate of $9\times 10^4$ s$^{-1}$ calculated from our measured value of $V_0$ and the detuning of the light. The effect of photon scattering on coherence will be the topic of future investigations.

In summary we have demonstrated a simple atomic interferometer that uses atoms, laser cooled into an optical lattice, followed  by an optical-lattice pulse. This technique is capable of producing atomic density modulations with a contrast significantly higher than the interferometer previously demonstrated in Ref. \cite{Cahn97}. This not only increases the signal-to-noise ratio of the interferometric signal, but could also be of interest in atomic lithography. In this field our technique has the advantage over the previously demonstrated optical mask technique  \cite{Turlapov05}
in that it generates an actual spatial atomic density modulation, and not a spatial internal state modulation. We investigated how the interferometer performs when a pulse violating the Raman-Nath condition is used and find that a small violation can improve the sensitivity, and increase the contrast of the atomic density modulation. For specific pulse lengths in this long-pulse interferometer we observe a coherent signal at times that differ greatly from the echo time. We showed that the interferometric signal can be used as a sensitive probe of the dynamics of the atoms in the optical lattice. We compared the experimental results to a simple 1D numerical calculation and found excellent agreement.

37.10.Jk
We gratefully acknowledge helpful discussions and comments from Bill Phillips, Kris Helmerson, Pierre Clad{\'{e}} and A. Tonyushkin and assistance in acquiring data from Clio Sleator. This work was supported by Office of Naval Research and NZ-FRST
contract NERF-UOOX0703. 

\end{document}